\documentstyle[11pt,paspconf,psfig]{article}

\newcommand{\etal}{{\it et al.}}

\newcommand{\msun}{M$_{\sun}$}

\begin{document}

\title{High Energy Continuum Spectra from X-Ray 
Binaries}

\vspace{-0.5cm}

\begin{center}

S.N. Zhang (zhang@ssl.msfc.nasa.gov)

{\it USRA/Marshall Space Flight Center, ES-84, Huntsville, AL 35812}

\end{center}

\vspace{-0.5cm}
\begin{abstract} A variety of high energy ($>$1 keV) spectra have been
observed in recent years from Black Hole (BH) and Neutron Star (NS)
X-ray Binaries (XB). Some common physical components exist between
BHXBs and NSXBs, resulting in some high energy spectral
features. A common component between a BHXB
and a weakly magnetized
NSXB is the inner accretion disk region extending very close to
the surface (for a NS) or the horizon (for a BH). The inner disk radiation 
can be described by a multi-color blackbody (MCB)
spectral model. The surface radiation of the NS can be
approximated by a Single Color Blackbody (SCB)
spectrum. For a strongly magnetized NSXB, the high energy emission
is from its magnetosphere, characterised by a 
thermal bremsstrahlung (TB) spectrum. 
In both BHXBs and weakly magnetized NSXBs, 
a hot electron
cloud may exist, producing the hard
X-ray power law (photon index -1.5 to -2.0) with thermal cutoff (50-200 keV).
It has been recently proposed that a converging flow may be formed near
the horizon of a BH, producing
a softer power law (photon index about -2.5) without cutoff up to several
hundred keV. Based on these concepts we also discuss possible ways to
distinguish between BH and NS XBs. Finally we discuss briefly spectral
state transitions in both BH and NS XBs.
\end{abstract}
\vspace{-0.5cm}

\section{Introduction} 

Following the launches of several
high energy astrophysical satellites in the last decade, 
for example, EXOSAT, TTM and HETE aboard Mir Space Station, ROSAT, {\it Ginga}
SIGMA/GRANAT, CGRO, ASCA, SAX and RXTE, our
knowledge of high energy spectra from XB sources has increased
enormously. In this review paper, we shall focus on the high energy
spectra of these binary systems whose primary electromagnetic energy
output is above 1 keV. We will, however, not discuss narrow or broad 
spectral line features.
\vspace{-0.2cm}
\section{Spectral formations in X-ray binaries}  

In a binary system involving a compact object (a NS or a BH) and a
more ordinary star (called companion star), matter can be transferred,
either through Roche lobe overflow or via stellar wind, from the
companion to the compact object. In order for the transferred or
accreted matter to reach the compact object, angular momentum 
loss has to occur, presumably via
the viscous mechanisms. Thus an accretion disk may be formed and the
gravitational energy loss via the viscous loss will be radiated in the
form of electromagnetic waves. The energy of the radiated photons will
become higher when the accreted material gets closer to the central
object. In the standard Shakura-Sunyaev (1973) model, a geometrically thin
and optically thick accretion disk is formed. The local radiation can
be approximated by a blackbody spectrum. The overall
spectrum is thus the integration of the local (effective) blackbody 
emission from the inner disk region to the outer disk edge (see e.g.,
Mitsuda \etal\ 1984, 1989). Such a spectrum model is called the
MCB model. The
majority of the total luminosity is radiated between the
inner disk edge (a few $R_{s}$, the Schwarzschild radius) 
to a radius of the order of 100 $R_{s}$. The inner edge temperature
for a stellar mass BH is usually around 1 keV. A MCB spectrum is shown in
figure 1 together with other spectral models to be discussed in this
section.

When the central compact object is a NS, the remaining gravitational
energy will be released when the accreted matter finally reaches the NS.
In the case of a weakly magnetized NS (B$\le$$10^{8}$ gauss), usually found in
atoll sources, 
the matter can reach the NS
surface directly and the radiated spectrum can be approximated by the
SCB model (1-2 keV) (Mitsuda \etal\ 1984),
although the boundary layer of the NS is quite complicated. 
The measured temperature is thus
the effective temperature of the NS surface. In the 1-10 keV region, the 
MCB spectrum is considerably softer
than the SCB spectrum. This is why the MCB
spectrum is usually
called an Ultra-Soft (US) spectrum.

In case of a strongly magnetized NS (B$\ge$$10^{10}$ gauss), usually
found in X-ray pulsars, the accreted matter cannot
reach the NS surface directly and is funneled onto its polar
cap region by the strong magnetic field. A hot plasma (5-20 keV) is
thus formed and the continuum emission can be described by the
TB, characterised by a hard power law (photon index
-2.0 to -1.0)
with an exponential cutoff below 20 keV (White, Swank \& Holt 1983). 
For a mildly magnetized NS ($10^{9}$$<$B$<$$10^{10}$ gauss), usually
found in Z-sources, both the SCB and the TB 
components may be present, so that the overall spectrum becomes quite
soft compared with X-ray pulsars. 

In some systems Roche lobe overflow cannot occur. Significant wind
accretion will take place if the companion can produce strong winds. Such
systems exist with either a NS (e.g., Vela X-1) or a BH (e.g., Cyg X-1)
as the compact object. As well as forming an accretion disk, the wind
accretion in a BH system may cause some interesting consequences. As
suggested by Shapiro and Lightman (1976), an asymmetric shock front (bow
shock) will be formed with a temperature of the order of 10$^{9}$ K,
when the BH accretes directly from the stellar
wind of its High Mass (HM) companion of Cyg X-1. It is reasonable to expect that
the thermal radiation from
the accretion disk, if formed, will be inverse Compton scattered to higher
energy photons (Zhang \etal\ 1997a) by the hot electrons in the bow shock
(the electrons are heated up by collisions with protons). 
As calculated by Sunyaev and Titarchuk (1980)
the spectrum will be characterized by a power law with a 
high energy cutoff, whose energy is proportional to the mean electron temperature in the
shock. This mechanism may naturally explain the observed hard state
spectrum of Cyg X-1 (photon index -1.5 to -2.0 with cutoff between 50-200 keV).
This is the so-called
TC process. 
In figure 2, we show an illustrative
model of Cygnus X-1 (Zhang \etal\ 1997a). 

Even in Low Mass (LM) XBs, some accretion disk models also suggest
the existence of a hot cloud surrounding the central compact object. 
Thus the TC process may also work here.
Several models are invoked to explain the formation of this
cloud. The most recent ones are the two-phase model of Haardt and Marashi
(1993), advection-dominated model of Narayan (1996) and the post-shock
model of Chakrabarti and Titarchuk (1995). Although this issue is still
controversial, this hot cloud seems to exist in both BH and NS XBs and
Seyfert galaxies as well, as inferred from the characteristic TC spectra. 

When the supply of the low energy photons is over-abundant, this
TC process can cool down the cloud and eventually destroy
it. In the case of a NS XB, the power law component will decrease and
finally disappear. For a BH XB, the situation is more interesting.
Because of the strong gravitational field and the absorptive horizon
of the BH, the low temperature electrons will move towards the BH
relativistically to form a converging flow (Chakrabarti and Titarchuk 1995). 
The relativistic electrons will up-scatter the low
energy photons, either trapped in the flow when it sweeps through the
surface of the accretion disk or simply intercepted by the flow, to
form a power law. We call this radiation mechanism Bulk Motion Comptonization
(BMC) process. 
Calculations made by Chakrabarti and Titarchuk
(1995) have shown that the BMC power law is steeper (photon index around -2.5)
than the TC
power law formed from the hot Comptonization cloud (photon index -1.5 to -2.0). 
\vspace{-0.2cm}
\section{Distinguishing between neutron star and black hole X-ray binaries}

As a matter of fact, up to the present time, the only reliable way of determining
if a binary system contains a BH is by measuring the system parameters.
If one can estimate the mass of the compact object,
it will be considered to be BH if its mass is
$>$ 3 \msun, currently the best known theoretical
and observational upper mass limit of a NS. Several systems have been
identified in this way. 
However, this method is quite restrictive in principally two aspects.
One is that the masses of the compact objects of the majority of the
BH candidates (see below) cannot be reliably measured due to various 
observational constraints.
The other is that this method in principle remains ambiguous for those
with mass limits $<$3 \msun. Therefore we have to look for alternative
methods of selecting BHBs.

By comparing the common characteristics of the high energy emission
from these dynamically determined BH systems, it has been hoped that some
``signatures" of BHs may be found when accurate
dynamical estimates are not available. Such
``signatures" include ultra-soft/ultra-hard energy spectra,
rapid variability, QPOs etc. All such features have, however, been
observed from NS systems, especially those with a weak magnetic field.
This is, perhaps, not surprising since most of these phenomena are
associated with properties of the accretion disk common to both types of 
systems. It has been recently
suggested that perhaps luminous hard X-ray emission or a combination
of an ultrasoft component and hard X-ray power-law are more indicative of the existence of
a black hole (van der Klis 1994; van der Klis and van
Paradijs 1993; White 1993; Tanaka \& Lewin 1995; Zhang \etal\ 1996a;
1996b; Barret, McClintoch \& Grindlay  1996).

In this section we try to understand how to distinguish between
NS and BH XBs, based on the similarities and differences in the
formation process of high energy spectra of these systems we discussed in
the previous section. We will first identify some signatures of NSXBs and then 
discuss how a BHXB might be identified in the absence of any signature
of NSXBs.

\subsection{Signatures of nuetron star X-ray binaries}

The high energy spectra of magnetized NSXBs can be relatively easily 
distinguished
from that of BHXBs. The TB spectral shape with
a cutoff energy below 20 keV is a strong indication of the existence
of the NS magenetosphere and, in fact, has never been detected from a BHXB.
The 1-2 keV SCB component, if unambiguously identified,
is a strong indication of the detection of the neutron star surface 
radiation. This component has also never been detected from a BHXB.
Thus both characteristic spectral components, together with pulsations and
type I X-ray bursts, can be used as signatures of NSXBs.
In the absence of any one of the above signatures, a source may be considered 
a BHC. We now discuss how to further identify those BHXBs from the list
of BHCs, such as that reviewed by Tanaka and Lewin (1995). 

\subsection{Indicators of a black hole X-ray binaries}

One common high energy radiating component between a BH and a weakly 
magnetized NS XB (or simply `NSXB' for the rest of this section since
the strongly magnetized NSXBs can be easily identified as discussed above), 
is the inner disk region very close to the compact
object, which produces the MCB spectrum. Due to the NS SCB emission, the overall
spectrum of such a NSXB may be harder than that of a BHXB.
However, when the mass accretion rate is low so that the NS surface
temperature is also low, the NSXB spectrum may appear indistinguishable from
that of a BHXB. {\it Thus a luminous ($>$10$^{37}$ erg/s) 
MCB component
is a strong indication of the existence of a BH in a XB.}

Using the MCB model, in principal one can
determine the temperature and radius of the inner disk edge. For 
a NSXB, 
the inner disk edge should be very close to the NS surface , i.e.,
about 10 km. For a BHXB, the inner disk radius can be much larger.
{\it Therefore an inner disk radius significantly larger than
10 km, for example, greater than 30 km, is another indication of
the existence of a BH.} This is why dynamically
determined BHXBs all appear to have much larger inner disk radii than that
of NSXBs (Ebisawa \etal\ 1991). However, for a maximally rotating Kerr BH with a mass
comparable to the NS mass, its inner disk radius will not be very different from
that of a NSXB. So a small inner disk radius is not always associated with a
NSXB. 
 
Another common high energy radiation component between a BHXB and a weakly
magetised NSXB is a hot electron cloud, producing the observed TC power 
law with cutoff between 50 and 200 keV. This cloud is different from
the extended corona in many the so-called `dipper' sources (The 
coronae in the `dipper' sources have much lower temperatures and extend
well above the plane of the disk, as inferred from the significant residual
fluxes of the high inclination systems during the accretion disk eclipse
by the companion.). Although such TC spectra are 
quite similar between NS and BH XBs, there are, however, still
some differences. One is that the spectral break energy in a BHXB is frequently
observed to be $>$ 100 keV, while for NSXBs the highest observed spectral
break energy is only 65 keV (Zhang \etal\ 1996a). Perhaps this is because 
there are two soft photon
sources in a NSXB (one is the disk emission and the other is the NS surface
emission), in contrast to the single soft photon source in a BHXB (the disk
emission). Therefore the electron temperature in a NSXB is usually lower
than that in a BHXB. The same argument can be applied to the fact that
the hard X-ray luminosity can be significantly higher in a BHXB. 
{\it Therefore the detection of a combination of a high energy cutoff
($>$100 keV) and a bright hard X-ray luminosity ($>$10$^{37}$
erg/s above 10 keV) indicates strongly the existence of a BH in the system.}  

The hard X-ray power law emission will disappear in a NSXB
when the disk accretion rate is so high that the hot electron cloud is
cooled down and consequently the TC mechanism can no longer effectively
produce hard X-ray photons. In a BHXB, a steeper power law (SPL) should be 
formed in
the converging flow by the BMC mechanism (Chakrabarti \& Titarchuk 1995). 
This emission
mechanism is related to the lack of a solid surface of a BH, and therefore may
provide a signature of the BH. The question is whether a similar high
energy spectrum can be also formed from other emission mechanisms not
related to the horizon of a BH. Although hypothetically many high
energy radiation mechanisms can form such a power law energy spectrum in 
this energy range (10-500 keV), the conditions when such a power law is 
detected restricts the possibilities severely. Almost in all cases 
the SPL (without cutoff up to 300-500 keV) is accompanied with 
a bright MCB component observed from a BHXB.

Synchrotron radiation is an unlikely mechanism, since
the bright MCB component tends to cool down the assumed
high energy
electrons by inverse Compton scattering, 
thus preventing non-thermal radiation up to several hundred keV. 
It has been widely considered
that the TC process with a cooler electron cloud as
the result of the more effective cooling by the higher soft X-ray
flux, may be 
a good candidate for producing the SPL. 
There are, however, several problems
with this picture. First, the lower electron temperature
should result in a lower spectral break energy when the SPL
is observed, contrary to the detected unbroken power law up to at least 300-500 
keV. The other problem is the near constancy of the SPL
index when both the MCB and the SPL luminosity change
significantly as observed from GS1124-68 in the soft state 
(Ebisawa \etal\ 1994). Other radiation mechanisms, involving relativistic
outflows or jets, are also unlikely to work, since not all BHXBs are observed with
jets or signficant radio emission, or such SPL
detection is not always associated with jet ejection events (Zhang \etal\
1996b). Moreover, all the above mentioned mechanisms should also operate
in a NSXB; but such SPL/MCB have never been detected from
any confirmed NSXB. {\it We therefore conclude that such unbroken
SPL (up to 300-500 keV) associated with a luminous 
MCB component ($\sim$10$^{37}$ erg/s) is mostly likely produced in the
converging flow near the BH horizon and may be used as a distinctive
signature of BHXBs.} 

In summary, there appears to be several strong suggestive indicators of
the existence of a BH in a XB. These are: a) a luminous ($\sim$10$^{37}$ erg/s)
MCB spectrum without the contamination of a SCB component; 
b) a bright ($\sim$10$^{37}$ erg/s) 
hard X-ray power law 
(photon index: -1.5 to -2.0) with high energy cutoff ($>$100 keV); and
c) a SPL (photon index $\sim$ -2.5) without 
cutoff up to 300-500 keV associated with a luminous 
MCB component ($\sim$10$^{37}$ erg/s). 
The first two indicators are related to the lack of surface radiation from
the BH, in contrast to a NS, while last one 
seems to be due to the horizon of the BH and thus is a more secured signature
of a BH. We do not include the large inner disk radius 
as a BH indicator here, since this parameter depends upon
the details of the rather complicated spectral model which is not well 
understood yet.

\subsection{Applications of black hole X-ray binary indicators}

As a matter of fact, none of the known NSXBs, including all known pulsars, 
Z-sources and type I X-ray bursters, have been observed with one of the three
high energy spectral types, with the only possible exception
of Terzan-2, a type I X-ray burster. During one observation with SIGMA/GRANAT, a hard power law
up to about 200 keV was detected with a luminosity of exceeding
$\sim$5$\times$10$^{37}$ erg/s (Barret \etal\ 1992). It is, however, not clear if the emission
originated from Terzan-2, or another high energy transient from that
direction or within the same globular cluster. At all other occasions
when it is detectable in the hard X-ray band above 20 keV, a much lower 
luminosity was seen. 

All the dynamically confirmd BHXBs, i.e., with the compact object mass
greater than 3 \msun, have displayed at least one of three indicators.
These systems are Cygnus~X-1 (indicator: b and c), 0620-00
(a, c),
LMC~X-3 (a, c), LMC~X-1 (a, c), GS~1124-683 (a, b, c), 
GS~2000+25 (a, c), GS~2023+33 (b), GRO~J1655-40 (a, c).
Here we also further examine all those remaining BHCs listed by Tanaka and Lewin (1995). These
systems should also contain a BH according to the three indicators: 
GX~339-4 (a, b, c), 1354-645 (c), 1543-475 (a, b), 1630-472 (a, b, c), 
1705-250 (a, c), 1740.7-2942 (b), 1741-322 (c), GRS~1915+105 (a) and
1758-258 (b). The most recently discovered transient GRS~1739-278
also belongs to this class since the indicator (c) was detected from it
(Vargas \etal\ 1996).

\section{Spectral state transitions in X-ray binaries}

We briefly summarise the spectral state transitions in BH and NS XBs, aiming
at understanding the dynamics in XBs.
 Here we call these states
`hard' or `soft', instead of `low' or `high' used by some others,
since we will show below that the bolometric luminosity,
presumably proportional to the overall mass transfer rate, actually
changes insignificantly or at least the hard state bolometric
luminosity is not necessarily always lower than that in the soft state.

\subsection{Transitions between the hard and soft states in Cyg X-1}

The most recent soft state of Cyg X-1 has been extensively observed
by several high energy instruments from aboard CGRO, RXTE, ASCA and SAX
satellites. Here we only show some results from the broad band
monitoring achieved with ASM/RXTE (1.3-12 keV) and BATSE/CGRO (20-200 keV)
(Zhang \etal\ 1997b). The main results are: a) during the whole
period the soft X-ray fluxes are anti-correlated with the hard X-ray fluxes;
b) the 1.3-200 keV luminosity (and the bolometric luminosity after
the color correction, see Zhang \etal\ 1997c for details) does not change
significantly, and is not correlated with the state transition; c)
the transition time of the lowest energy band (1.3-3.0), whose flux is dominated
by the MCB component, is much shorter than
that of the hard X-ray flux; d) flares with durations of around 10-30 days
are seen in all energy bands, superposed on top of the state transition;
and e) the soft X-ray flares lag behind the hard X-ray flares by 10-20 days.

The near constant luminosity during the state transitions defies most of the
current models of high energy radiation in BHXBs. All these models require
a significant luminosity increase in the soft state, although some fine tuning
of some of these models may be able to remove this requirement. For example,
in the Chakarabarti and Titarchuk (1995) model, at a certain mass accretion rate
the re-distribution of the mass flow rates between the Keplerian and sub-Keplerian
mass accretion components may also produce a constant luminosity during a state transition
(Titarchuk 1996), instead of requiring a total mass acrretion rate
increase. However, the other results, especially the longer hard X-ray
flux transition time and the lag between the soft and hard flares cannot be
naturally accounted for within these current models.

\subsection{State transitions in GS~1124-683 = Nova Muscae}

There appears to be several spectral state transitions
observed from the LM BHXB GS~1124-683 during its 1991 outburst
(see Ebisawa \etal\ for details of this outburst observed with
{\it Ginga}). Here we just briefly mention two behaviors. The first one
is during the initial outbursting phase, when the power law component 
reached its peak flux about 3 days earlier than the MCB
component. This behavior is not really a spectral state transition,
but instead reflected a lag of the MCB component increase
behind the power law component increase, in responding perhaps to the
mass accretion rate increase. It is interesting to note that when
the soft component reached its peak, the hard component also experienced
a dip, without significant changes of the spectral shape of each component.
The second one is a real spectral state transition between April and July,
when the hard component reached a secondary maximum and the soft component
continued to decrease. The significant features of this state transition
are the low luminosity level ($\sim$1\% of the peak total luminosity) when
the transition happened and the much harder power law (photon index
changed from between -2.2 and -2.7 to $\sim$ -1.6) during this
secondary outburst.

\subsection{Soft to hard state transition in 4U~1608-522} 

In figure 3 we show the spectral transition of the LM NSXB 4U~1608-522 when the
energy spectrum changed from a typical soft state spectrum 
(MCB plus a SCB) to a hard state 
(power-law dominated) spectrum when the total
luminosity decreased from about 5$\times$10$^{\rm 37}$ to about 
7$\times$10$^{\rm 36}$ erg/sec. A transition happened at
10$^{37}$ erg/sec. At this point the MCB component (F$_{\rm BD}$)
decreased by nearly a factor of 3, the hard component (F$_{\rm C}$)
(believed due to TC in an optically thin region) increased by
about a factor of 2 and the SCB (F$_{\rm NS}$) remained
almost constant or increased slightly. The total X-ray luminosity
tracks the
total mass accretion in the system since the majority of the
gravitational energy is released in the X-ray energy band. Therefore
during this transition the change in total mass accretion rate is very
small. This is also indicated by the near constant SCB component
during the spectral state transition. 
The sudden decrease of the MCB component indicates the mass
accretion rate through the inner edge of the disk decreased
by about a factor of 3. These material must be channeled to
the region producing the observered hard X-ray component which increased
simultaneously.

\subsection{Possible Interpretations}

 A successful model has to account
for several major phenomena. These are: {\it a)} the soft X-ray component lags 
behind the hard component
during the flares or onsets of outbursts (hard precursors are common
among soft X-rat transients or X-ray novae);
b) no significant luminosity changes occur during state 
transitions (the spectral state transitions are usually shown
as soft/hard component anti-correlations. After removing the trend of long term
light curve exponential decay during the GS~1124-68 spectral state transition,
the total luminosity changes are also not significant.); and c) the state
transition time of the soft component is much shorter than that of the hard
component (probably only observed from Cyg~X-1, since no other 
sources have been monitored continuously over the whole transition period
in a broad energy band). Since all three phenomena are observed from Cyg~X-1,
we will examine this system first.

In the model we show in figure 2 for Cyg~X-1, phenomenon-{\it a} can be explained
as due to the time difference between the disk mass transfer and the 
direct wind capture by the BH. When the mass flow rate carried by the stellar 
wind
increases suddenly, the flow will reach the bow shock promptly, resulting
in an increased hard X-ray flux, due to the increased Comptonization
factor, although the supply of the seed photons remains the same. The mass 
accretion component transported 
through the optically thick disk will arrive later at the inner disk region,
thus producing the delayed
soft X-ray flux. The delay time of around 15 days constrains the optically thick 
Keplerian disk to fill only 1-10\% of the Roche lobe
of the system (Zhang \etal\ 1997a). Phenomenon-{\it b} suggests that the
the total mass accretion rate remains nearly a constant while
both soft and hard X-ray fluxes change significantly. It is therefore
reasonable to assume what really happened was the re-distribution of mass flow
between the direct wind capture component and the optically thick disk
component. Phenomenon-c is explained as due to the mass flow exchange between the
two mass accretion components when an instability is transported between
the outer and inner disk regions (see Zhang \etal\ 1997a for details).

Although no significant wind accretion exists in LMXBs, many models will
require the existence of an optically thin, sub-Keplerian region, responsible
for the hard X-ray production, although the origin and the nature of this
region distinguishes these models. In this picture, the optically thick disk
nearly
fills the Roche lobe, since it is formed out of the Roche lobe overflow, and
the optically thin region surrounds the inner disk region nearly spherically.
The time lag is then the same time difference, but starting from the boundary
of the optically thin region since its radius is expected to be smaller than
the outer radius of the optically thick disk. {\it Thus the time lag measures
the size of the optically thin region in a LMXB.}
In GS~1124-68, the lag time of $\sim$3 days
suggests the size of this region is between 10$^{9}$ to 10$^{10}$ cm, i.e,
about 10-100 times larger than the inner disk radius for a stellar mass BH.
During the Cen~X-4 outburst in 1979 (Bouchacourt \etal\ 1984), 
a similar lag was also found, indicating a similar geometry in both systems.
A significant difference between Cyg~X-1 and these LMXBs, in terms of 
spectral state transitions, is that the hard state is the persistent 
and dominating state and occurs at a much higher luminosity level in Cyg~X-1
than the LMXBs. This is, perhaps due to the relatively higher mass flow
rates in the optically thin region originated from the direct stellar
wind capture in our model of Cyg~X-1, than that in LMXBs, where the optically
thin region is probably formed out of the optically thick disk and is much
less effective when the disk mass accretion rate is very high.

\section{Summary}

It appears that a
characteristic high energy spectral component is formed from each component
of a NS or BH XB. Based on the similarities and differences of these spectral
formation process in NS and BH XBs, we discussed possible ways to distinguish
between NS and BH XBs. Spectral state transitions in XBs are discussed for
three systems, HMBH XB Cyg~X-1, LMBH XB GS~1124-68 and LMNS XB 4U~1608-522. 
Geometries of these systems can be constrained from the observed spectral
transition characteristics.

\acknowledgements

We benefitted greatly from the many stimulating
discussions between the BATSE team members, especially Alan Harmon, Bill
Paciesas and Craig Robinson at 
Marshall Space Flight Center
and with Drs. Ken Ebisawa, Lev Titarchuk, Sandip Chakrabarti, Wei Cui,
Wan Chen, Jan van Paradijs, Didier Barret, Ron Remillard, Bob Hjellming 
and Ramesh Narayan.

\begin{figure}
\hbox{
\vbox{
\psfig{figure=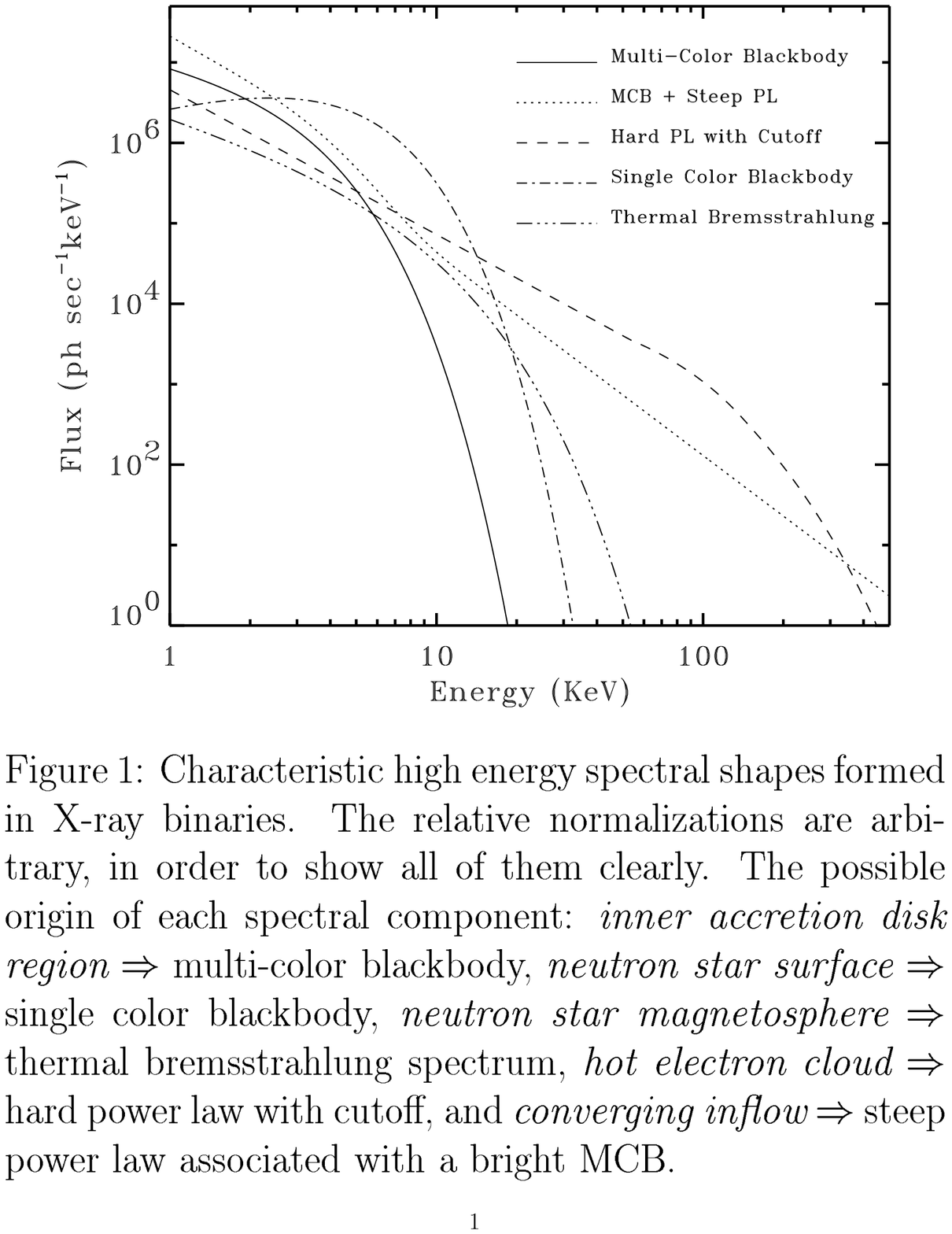,width=3.3in,%
bbllx=40bp,bblly=110bp,bburx=530bp,bbury=720bp,%
clip=}
\psfig{figure=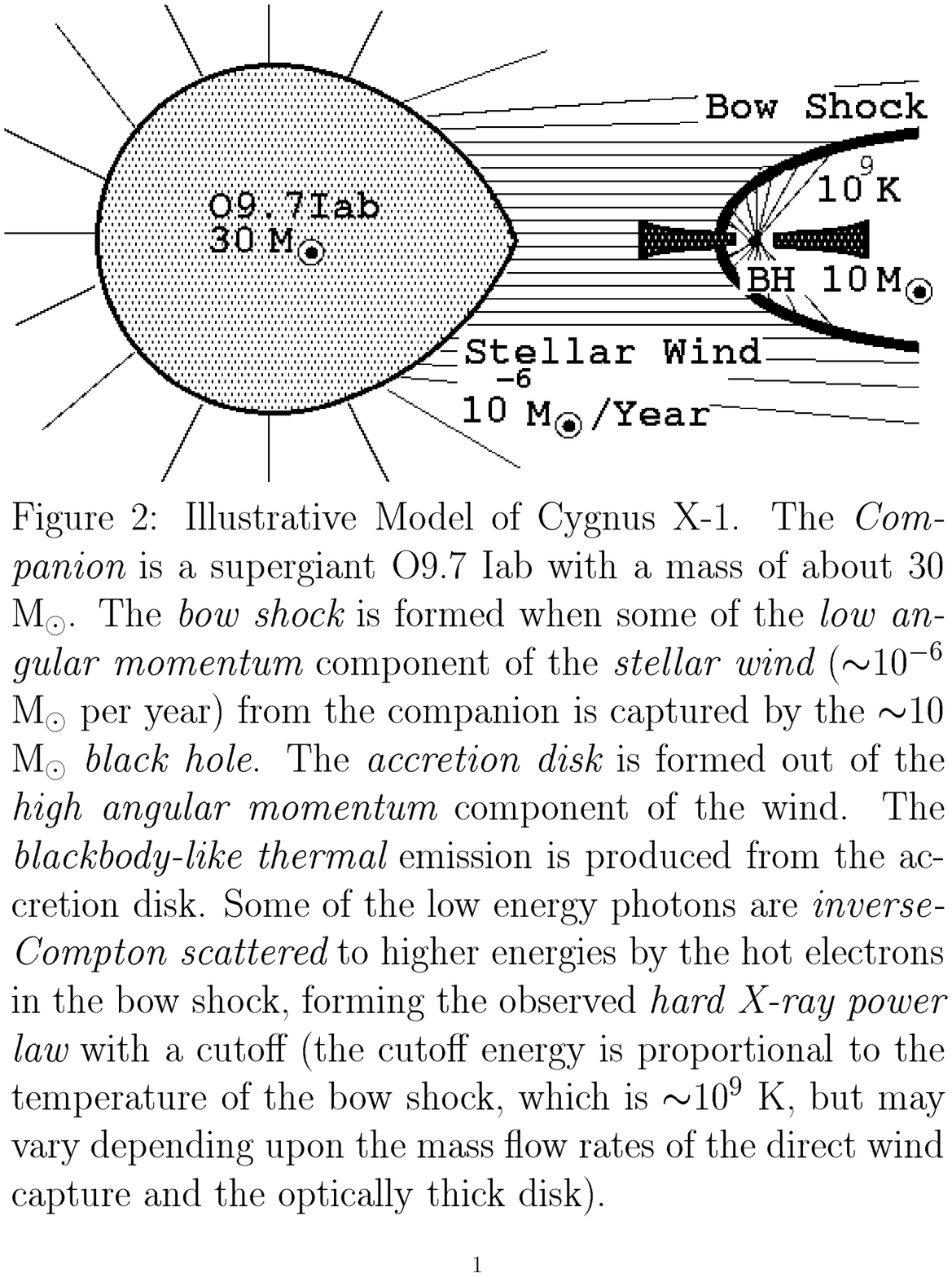,width=3.3in,%
bbllx=40bp,bblly=108bp,bburx=530bp,bbury=740bp,%
clip=}
}
\vbox{
\psfig{figure=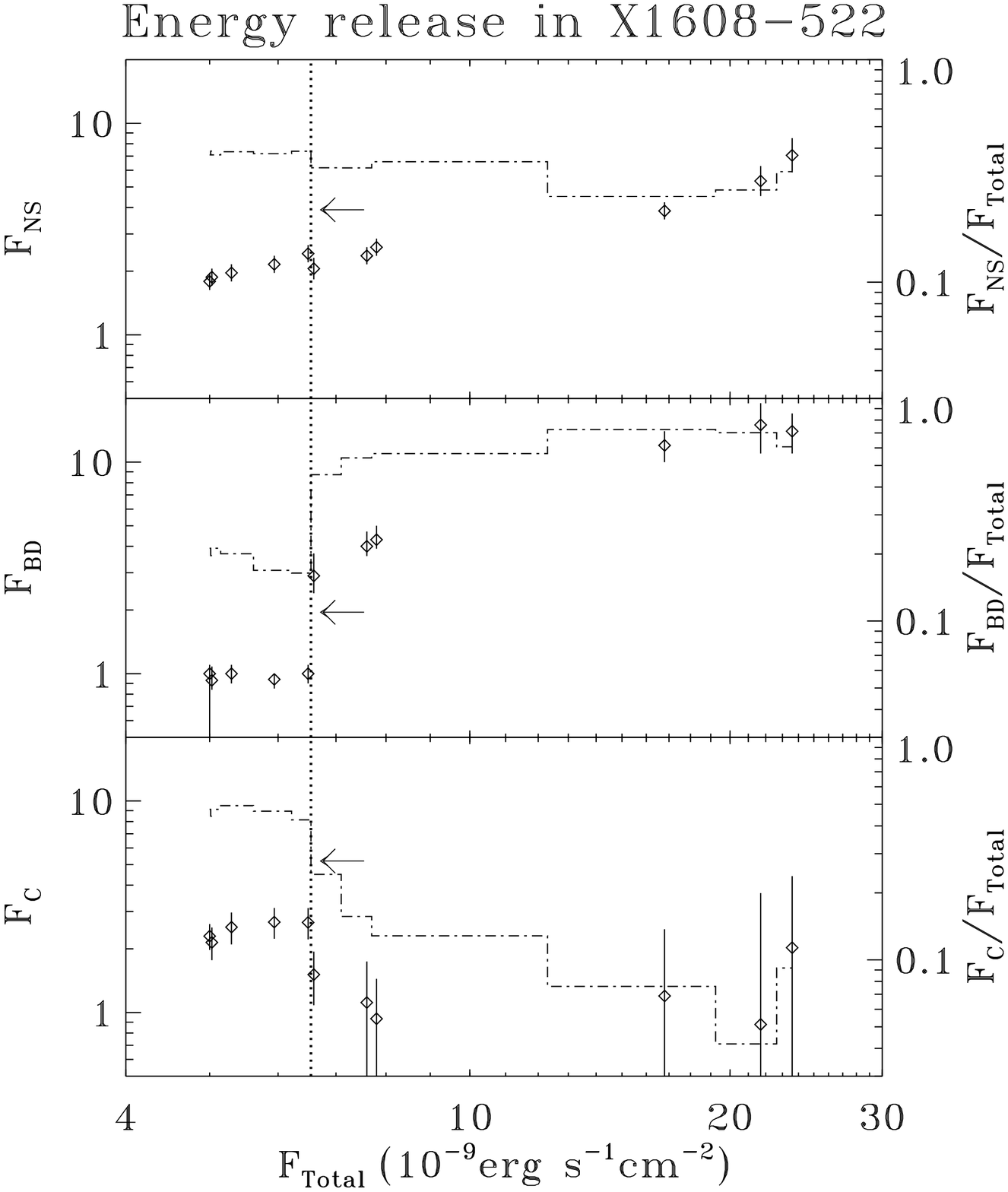,width=2.0in}
\vspace{0.5cm}
\parbox[b]{1.6in}{\small Figure~3: Energy spectral
state transition observed from 4U~1608-522 with {\it Tenma}.
F$_{\rm total}$
is the total detected energy flux, which can be further divided into
three components. F$_{\rm NS}$ is the SCB emission
from the NS surface, F$_{\rm BD}$ is the MCB Component
from the inner region of the optically thick disk and F$_{\rm C}$ is the energy flux added by
a Comptonization process (all in unit of 10$^{-9}$ erg s$^{-1}$ cm$^{-2}$).
Diamonds denotes the energy fluxes and the {\it dot-dashed} lines are
the percentages of these three components within the total observed
energy flux.
A transition happened when the total luminosity was about 6.5$\times$
10$^{-9}$ erg s$^{-1}$ cm$^{-2}$, corresponding to $\sim$ 10$^{37}$ erg s$^{-1}$ for
at $\sim$3.6 kpc. While
the total luminosity did not change significantly, the gravitational
energy release was mostly transferred from the optically thick disk to the
Comptonization region, possibly the optically thin mass accretion 
region (Data from {\it Tenma} observations described by
Mitsuda {\it et al.} 1989).}
}
}
\end{figure}
\end{document}